\documentclass[twocolumn,10pt,amsmath,amssymb,pra,aps,superscriptaddress, nofootinbib, longbibliography]{revtex4-1}
\usepackage[utf8]{inputenc}
\usepackage{graphicx}
\usepackage{hyperref}
\graphicspath{ {imgs/} }

\usepackage{xcolor}
\usepackage{booktabs}
\usepackage{bm}

\newcommand{\eF}{\varepsilon_{F}}

\newcommand{\kF}{k_{\textrm{F}}}
\newcommand{\xibcs}{\xi_\mathrm{BCS}}

\begin{document}

\title{Vortex lattice in spin-imbalanced unitary Fermi gas}

\author{Jakub Kopyci\'{n}ski}
\affiliation{Faculty of Physics, Warsaw University of Technology, Ulica Koszykowa 75, 00-662 Warsaw, Poland}
\affiliation{Center for Theoretical Physics, Polish Academy of Sciences, Al. Lotnik\'{o}w 32/46, 02-668 Warsaw, Poland}

\author{Wojciech R. Pude\l{}ko}
\affiliation{Faculty of Physics, Warsaw University of Technology, Ulica Koszykowa 75, 00-662 Warsaw, Poland}
\affiliation{Swiss Light Source, Paul Scherrer Institut, CH-5232 Villigen PSI, Switzerland}
\affiliation{Physik-Institut, Universit\"at Z\"urich, Winterthurerstrasse 190, CH-8057 Z\"urich, Switzerland}

\author{Gabriel Wlaz\l{}owski}\email{gabriel.wlazlowski@pw.edu.pl}
\affiliation{Faculty of Physics, Warsaw University of Technology, Ulica Koszykowa 75, 00-662 Warsaw, Poland}
\affiliation{Department of Physics, University of Washington, Seattle, Washington 98195--1560, USA}

\begin{abstract}
We investigate the properties of a spin-imbalanced and rotating unitary Fermi gas. Using a density functional theory (DFT), we provide insight into states that emerge from a competition between Abrikosov lattice formation, spatial phase separation, and the emergence of  Fulde-Ferrell-Larkin-Ovchinnikov (FFLO) state. A confrontation of the experimental data [M. Zwierlein {\it et al.}, Science {\bf 311}, 492 (2006)] with theoretical predictions provides a remarkable qualitative agreement. In the case of gas confined in a harmonic trap, the phase separation into a superfluid core populated by the Abrikosov lattice and a spin-polarized corona is the dominant process. Changing confinement to a box-like trap reverts the spatial location of the component: gas being in the normal state is surrounded by superfluid threaded by quantum vortices. The vortex lattice no longer exhibits the triangular symmetry, and the emergence of exotic geometries may be an indirect signature of the FFLO-like state formation in the system.
\end{abstract}

\maketitle

\section{Introduction}

The most fundamental technique for probing the superfluid properties of a system is via its response to rotation. This technique was used to demonstrate the superfluid character of a Bose-Einstein condensate by direct observation of the Abrikosov vortex lattice~\cite{Abo-Shaeer476}. Later on, the same technique was applied to ultracold fermionic gases, providing direct evidence for the occurrence of the superfluidity over the entire BCS-BEC crossover~\cite{Zwierlein2005}. Immediately, the focus was shifted towards the spin-imbalanced Fermi gas, where the principal mechanism responsible for the superfluidity, namely Cooper pairing, is modified. Theoretically, in spin-imbalanced systems Fulde-Ferrell-Larkin-Ovchinnikov (FFLO) state should emerge~\cite{Takahashi2006,Shim2006,Bausmerth2008,Kulic2009,Shang2010,Warringa2012}. The FFLO state corresponds to a ground state of the spin-imbalanced system, where the pairing between opposite spin partners with momentum $\bm{p}$ and $\bm{-p}+\bm{q}$ takes place. It manifests as the order parameter fluctuations in form of $\Delta(\bm{r})\sim e^{i\bm{q}\cdot\bm{r}}$ (FF) or $\Delta(\bm{r})\sim \cos(\bm{q}\cdot\bm{r})$ (LO). In practical realization, these oscillations patterns are modified by the geometrical effects. In the case of experiments with ultracold atomic gases, the primary source of modifications comes from the trapping potential, see~\cite{GUBBELS2013255,Radzihovsky_2010,Kinnunen2018} for extensive overviews. 
In general, the trapping effects limit the FFLO state to be confined only in a limited volume. Moreover, it competes with the phase-separation phenomenon~\cite{Partridge2006,Partridge2006,Shin2006,Liao2010,Kim2011,Olsen2015}, i.e. the situation where the system separates spontaneously into a spin-symmetric superfluid part (so-called BCS state) and a fully polarized state. In the case of the most popular harmonic traps, the BCS superfluid occupies the trap center while the spatial fluctuations of the order parameter take place close to the system's boundary. 

Recently, it has been predicted that the presence of the quantum vortices may stabilize the FFLO state in a wide range of control parameters~\cite{Inotani2021}. This is because the vortices naturally introduce singularity lines $\Delta=0$ to the system. The population imbalance can accommodate there~\cite{Takahashi2006,Hu2007,2011.13021}. Thus, locally around the line, the system can support conditions needed for the realization of the FFLO state, which in this case should manifest as the order parameter modulation as a function of distance from the vortex core. Moreover, the standard technique of generating vortices is due to the rotation of the trap. Then, the vortices localize typically close to the rotation axis, typically located in the trap center. In this way, the problem of the FFLO state generation close to the boundary can be suppressed. This simple picture is derived from studies of a single quantum vortex accompanied by the FFLO state. Unfortunately, presently available experimental measurements of the rotating spin-imbalanced Fermi gas~\cite{Zwierlein_2006} have not provided a clear signature of the FFLO state formation. Also, no indications of vortices with exotic internal structures were observed.

Superconductors, typically two-dimensional, are another class of materials where FFLO state can be realized; see review paper~\cite{Croitoru2017}. In these systems, the FFLO state can be induced by applying an external magnetic field, which naturally favors the emergence of Abrikosov vortex states as well.
These vortex states couple with the FFLO and make it difficult to identify them unambiguously. The difficulties in the interpretation triggered many studies targeted at a better understating of the coexistence of vortex lattice states and FFLO state.  The studies include considerations of setups with single vortex line~\cite{Mizushima2005,Ichioka2007,Ichioka2007-2,Shimahara2009,Yokoyama2010} as well as the setups with vortex lattices~\cite{Shimahara1997,Klein2000,Yang2004,Klein2004,Sugiura2019,Suzuki2020}. Due to energetic reasons, the vortices tend to form triangular structures. At the same time, the nodal lines in the FFLO state prefer other geometries, where the most natural are square lattices emerging for $\Delta(x,y)\sim \cos(q_x x)\cos(q_y y)$. The appearance of the vortex lattice of an exotic structure may be regarded as an indirect signature of the FFLO state emergence. This type of indirect protocol has also been suggested in the context of ultracold atomic gases~\cite{Shim2006}. 

In this paper, we provide results of numerical calculations for spin-imbalanced Fermi gas being the subject of rotation by means of a density functional theory (DFT)~\cite{Bulgac_2012}. It is a fully microscopic approach, designed for dealing with strongly interacting systems, with an explicit treatment of fermionic degrees of freedom. The rotating cloud of the unitary Fermi gas (UFG) is the subject of our studies. The interaction is tuned to the resonance $a\kF\rightarrow\infty$ where $a$ is the scattering length and $\kF=(3\pi^2 n)^{1/3}$ is the Fermi wave vector corresponding to the density $n$. The choice of the investigated system is motivated by the availability of the experimental data~\cite{Zwierlein_2006}, which serves here as a reference point for comparing with theoretical predictions. The uniform unitary gas, similarly to the free Fermi gas, is characterized only by a single scale: average interparticle distance $n^{1/3}\sim\kF$. This property
significantly limits the possible forms of an energy density functional (EDF), and it allowed for the successful formulation of the DFT framework for the UFG -- the approach being in principle exact, see for review~\cite{Bulgac_2012}. The framework has been already applied for studies of quantum vortices~\cite{Bulgac2003,Bulgac2011,Bulgac_2014,Wlazowski2015}, revealing a remarkable agreement with experiment~\cite{Wlazlowski_2018}. Also, it admits the development of the FFLO state at unitarity~\cite{Bulgac2008}.  Applying the DFT framework to the system that combines vortices and spin-polarization effects will serve as the next step of the validation process of the theory. Note that the spin-imbalance effects are beyond the reach of simplified methods that were used so far for studies of vortex lattices, like Gross-Pitaevskii equation or  Local Phase Density Approximation to the Bogoliubov-de Gennes~ equations~\cite{Simonucci2015}. In the Bose-Einstein condensate limit it is possible to describe the system by set of coupled equations for composite bosons and excess fermions~\cite{PhysRevLett.96.150404}.

These studies are also relevant for neutron stars, particularly for magnetars~\cite{Turolla2015,Blaschke2018}. They are assumed to be superfluid and filled with a large number of vortices. Observed rapid change of rotation frequency, called neutron star glitch,  is regarded as a
direct manifestation of the superfluid character of the star
interior~\cite{BAYM1969,ANDERSON1975,Pines1985,Haskell2015}.
Moreover, a strong magnetic field of the order or larger than $10^{16}$~G is expected to be present inside the magnetars. It can effectively spin polarize the neutron matter and induce unconventional superfluid phases, including FFLO~\cite{Stein2012,Stein2014,Stein2016}. Whether the FFLO state can develop in the rotating star remains an unanswered question. The only methodology that can provide insight into this problem is through numerical simulation. Presently, the DFT method has become one of the standard tools for microscopic studies of nuclear systems as well~\cite{RevModPhys.75.121,Bulgac2019,Col2020}, and in principle, it can also be applied to rotating and spin-polarized neutron matter. However, a prior validation of the framework by applying it to strongly interacting terrestrial systems under similar conditions is desired. The ultracold atomic setup, as discussed here, suits this purpose very well.

\section{Framework}
In the calculations, we use the density functional theory. The chosen energy density functional is known as asymmetric superfluid local density approximation (ASLDA), designed specifically for the strongly interacting Fermi gas at unitarity~\cite{Bulgac_2012}. It has the following form (we use metric system $m=\hbar=k_B=1$):
\begin{equation}
\begin{split}
\mathcal{E}_{\mathrm{aslda}}=\sum_{\sigma=\{\uparrow, \downarrow\}}\frac{\alpha_\sigma(p) \tau_\sigma}{2}+\beta(p)(n_{\uparrow}+n_{\downarrow})^\frac{5}{3}+ \\ +\gamma(p)\frac{\nu^\dag \nu}{(n_{\uparrow}+n_{\downarrow})^{\frac{1}{3}}}+\sum_{\sigma=\{\uparrow, \downarrow\}}\left[1-\alpha_\sigma(p)\right]\frac{\bm{j}_\sigma^2}{2 n_\sigma}.
\end{split}
\label{eq:ASLDA}
\end{equation}
This functional consists of the kinetic term, normal and pairing interaction terms and a term responsible for restoring the Galilean invariance, respectively. These are functions of particle density $n_{\sigma}$, the kinetic density $\tau_{\sigma}$, the anomalous density  $\nu$ and the probability current $\bm{j}_{\sigma}$, where $\sigma=\{\uparrow, \downarrow\}$ indicates the spin components. The densities and currents are parametrized via set of orthonormal set of Bogoliobov quasiparticle wave functions $\{u_{n,\uparrow}(\bm{r}),u_{n,\downarrow}(\bm{r}),v_{n,\uparrow}(\bm{r}),v_{n,\downarrow}(\bm{r})\}$ as follow
\begin{eqnarray}
n_\sigma(\bm{r})&=&\sum_{|E_n|<E_c} |v_{n,\sigma}(\bm{r})|^2f_{\beta}(-E_n),\\
\tau_\sigma(\bm{r})&=&\sum_{|E_n|<E_c} |\bm{\nabla} v_{n,\sigma}(\bm{r})|^2f_{\beta}(-E_n),\\
\nu(\bm{r}) &=&\sum_{|E_n|<E_c} u_{n,\uparrow}(\bm{r})v_{n,\downarrow}^{*}(\bm{r})\frac{ f_{\beta}(-E_n)-f_{\beta}(E_n)}{2},\\
\bm{j}_\sigma(\bm{r})&=&\sum_{|E_n|<E_c}\Im\left(v_{n, \sigma}(\bm{r})\bm{\nabla} v^*_{n, \sigma}(\bm{r})\right)f_{\beta}(-E_n).
\end{eqnarray}
These wave functions have the interpretation that $|u_{n, \sigma}(\bm{r})|^2$ is the probability density of the $n$-th state being occupied with a spin $\sigma$ hole and $|v_{n, \sigma}(\bm{r})|^2$ with a spin $\sigma$ particle. The Fermi-Dirac distribution $f_{\beta}(E)=1/(\exp(\beta E)+1)$ is introduced in order to model temperature $k_B T=\beta^{-1}$ effects. To avoid divergences, only states with quasiparticle energies smaller than the cutoff energy $E_c$ are considered and regularization scheme as described in~\cite{Bulgac2002r,Bulgac_2012} is applied. The coupling constants $\alpha_\sigma(p)$, $\beta(p)$ and $\gamma_\sigma(p)$ are functions of the local spin polarization of the system $p(\bm{r})=\frac{n_{\uparrow}(\bm{r})-n_{\downarrow}(\bm{r})}{n_{\uparrow}(\bm{r})+n_{\downarrow}(\bm{r})}$. They are represented as polynomials, and fitted to  the quantum Monte Carlo calculations for spin-symmetric as well as spin-imbalanced systems~\cite{Bulgac_2012}.

Equations of motion are determined by minimization of the functional
\begin{equation}
\begin{split}
E=\int \left[ \mathcal{E}_{\textrm{aslda}}(\bm{r})+V_{\textrm{ext}}(\bm{r})n(\bm{r})\right]d\bm{r}\\
-\mu_{\uparrow}N_{\uparrow}-\mu_{\downarrow}N_{\downarrow}-\Omega_z L_z,
\end{split}
\end{equation}
where $V_{\textrm{ext}}$ is the external potential that couples to the total density $n=n_{\uparrow}+n_{\downarrow}$, $\mu_\sigma$ are chemical potentials (Lagrange multipliers) that control particle number $N_{\sigma}=\int n_\sigma(\bm{r})\,d\bm{r}$, and $\Omega_z$ is angular frequency that couples to $z$-th component of the angular momentum $\bm{L} = \int \bm{r}\times(\bm{j}_{\uparrow}+\bm{j}_{\downarrow})\,d\bm{r}$. The last term is equivalent to the transformation of the problem to a rotating frame. 

The minimization with respect to the quasiparticle orbitals provides equations that formally have structure of Bogoliubov-de Gennes equations:
\begin{equation}
\begin{split}
\begin{pmatrix}
h_{\uparrow}-\mu_{\uparrow}-\Omega_z l_z & \Delta\\
\Delta^* & -\left(h_{\downarrow}-\mu_\downarrow-\Omega_z l_z\right)^*
\end{pmatrix}
\begin{pmatrix}
u_{n, \uparrow}\\
v_{n, \downarrow}
\end{pmatrix}=\\=E_n\begin{pmatrix}
u_{n, \uparrow}\\
v_{n, \downarrow}
\end{pmatrix},
\end{split}
\label{eq:bdg}
\end{equation}
with $h$ being the single-particle Hamiltionian 
\begin{equation}
h_{\sigma} = -\dfrac{1}{2}\bm{\nabla}\alpha_{\sigma}\bm{\nabla} + \dfrac{\delta\mathcal{E}_{\textrm{aslda}}}{\delta n_{\sigma}} + V_{\textrm{ext}}-\dfrac{i}{2}\left\lbrace \dfrac{\delta\mathcal{E}_{\textrm{aslda}}}{\delta \bm{j}_{\sigma}},\bm{\nabla} \right\rbrace
\label{eq:hsigma}
\end{equation}
and $\Delta=-\frac{\delta\mathcal{E}_{\textrm{aslda}}}{\delta \nu^*}$ being the pairing field, which plays a role of the superfluid order parameter. Single particle angular momentum operator is $l_z=-i\left(x\frac{\partial}{\partial y}-y\frac{\partial}{\partial x}\right)$. The remaining components of the quasiparticle wave functions $u_{n, \uparrow}$ and $u_{n, \downarrow}$ are obtained by using the symmetry relation: $u_{n, \uparrow}\mapsto v^*_{n,\uparrow}$, $v_{n, \downarrow}\mapsto u^*_{n,\downarrow}$ and $E_n\mapsto -E_n$.

The simulations consider the following scenarios: the non-rotating gas trapped in a harmonic potential and the rotating gas in a harmonic and in a box trap. In all cases, we assume that the trapping acts only in $x$ and $y$ directions, while in $z$ direction system is uniform. 
For example, the harmonic potential has the form of $V_\mathrm{ho}(\bm{r})=\frac{1}{2}m\omega^2\rho^2$, where $\rho=\sqrt{x^2+y^2}$.
This simplification allows us to consider systems with a number of vortices $\sim 50$, which is the same order as observed in the experiment.  
In the case of the box potential, we assumed that the system is kept in a tube of radius $R_t$. Precisely, the potential is zero for $\rho<R_t$, and next within interval $\Delta R_t\ll R_t$ it smoothly rises to constant value $4\eF$, where $\eF=\kF^2/2$ is the Fermi energy. 
In each scenario, we keep the total number of atoms $N=N_\uparrow+N_\downarrow$ fixed, and we vary the spin imbalance $\delta=(N_\uparrow-N_\downarrow)/N$. The set of considered imbalances is the following: $\delta=0, 5\footnote{For the gas in the box potential only.}, 10, 18, 28, 39, 50, 62, 80$ and $90\%$.

The simulations were executed on a three-dimensional spatial
mesh of size $N_x \times N_y \times N_z$, and periodic boundary conditions are assumed. The lattice spacing was selected to be $dx\approx \xibcs$, where $\xibcs=\kF/\pi\Delta$ is the BCS coherence length, that defines expected vortex size. Here, as the Fermi momentum $\kF=(3\pi^2n)^{1/3}$ and the pairing field $\Delta$ we used values corresponding to the center of the trap. The system temperature is close to the absolute zero and reads  $k_BT=0.01\eF$. The angular frequency of the harmonic trap was set to $\omega/\eF=33.7\cdot 10^{-3}$, and $R_t=56\xibcs$ for the box potential. When considering the rotating systems, we used $\omega/\Omega_z=\frac{110}{70}$, which matches the ratio used in the experiment~\cite{Zwierlein_2006}. The simulation parameters are summarized in Tab.~\ref{tab:params}. The Eqs~(\ref{eq:bdg}) were solved self-consistently, and the Broyden algorithm was applied to improve the convergence rate~\cite{Baran2008,Johnson1988}. The calculations were executed with the use of W-SLDA Toolkit~\cite{Wlazlowski_2018, Bulgac_2014, WSLDATookit}

\begin{table}[h]
    \caption{\label{tab:params}
    Parameters used in simulations: number of atoms $N$, harmonic trap angular frequency $\omega$, rotation angular frequency $\Omega_z$, lattice size and spacing $dx$. 
    }
    \begin{ruledtabular}
    \begin{tabular}{ccccc}
     $N$ & $\omega/\eF$ & $\Omega_z/\eF$ & Lattice size & $dx/\xibcs$\\
     \hline
     \multicolumn{5}{c}{Non-rotating case in the harmonic potential}\\
     \hline
    $16324$ & $33.7\cdot10^{-3}$ & $0$ & $96^2 \times 48$ & $1.33$\\
    \hline
     \multicolumn{5}{c}{Rotating case in the harmonic potential}\\
     \hline
    $8162$ & $33.7\cdot10^{-3}$  & $21.7\cdot10^{-3}$ & $128^2 \times 32$ & $1$\\
     \hline
     \multicolumn{5}{c}{Rotating case in the box potential}\\
     \hline
    $34032$ & N/A &  $21.7\cdot10^{-3}$ & $128^2 \times 32$ & $1$
    \end{tabular}
    \end{ruledtabular}
\end{table}

\begin{figure*}[t]
    \centering
    \includegraphics[width=0.83\linewidth]{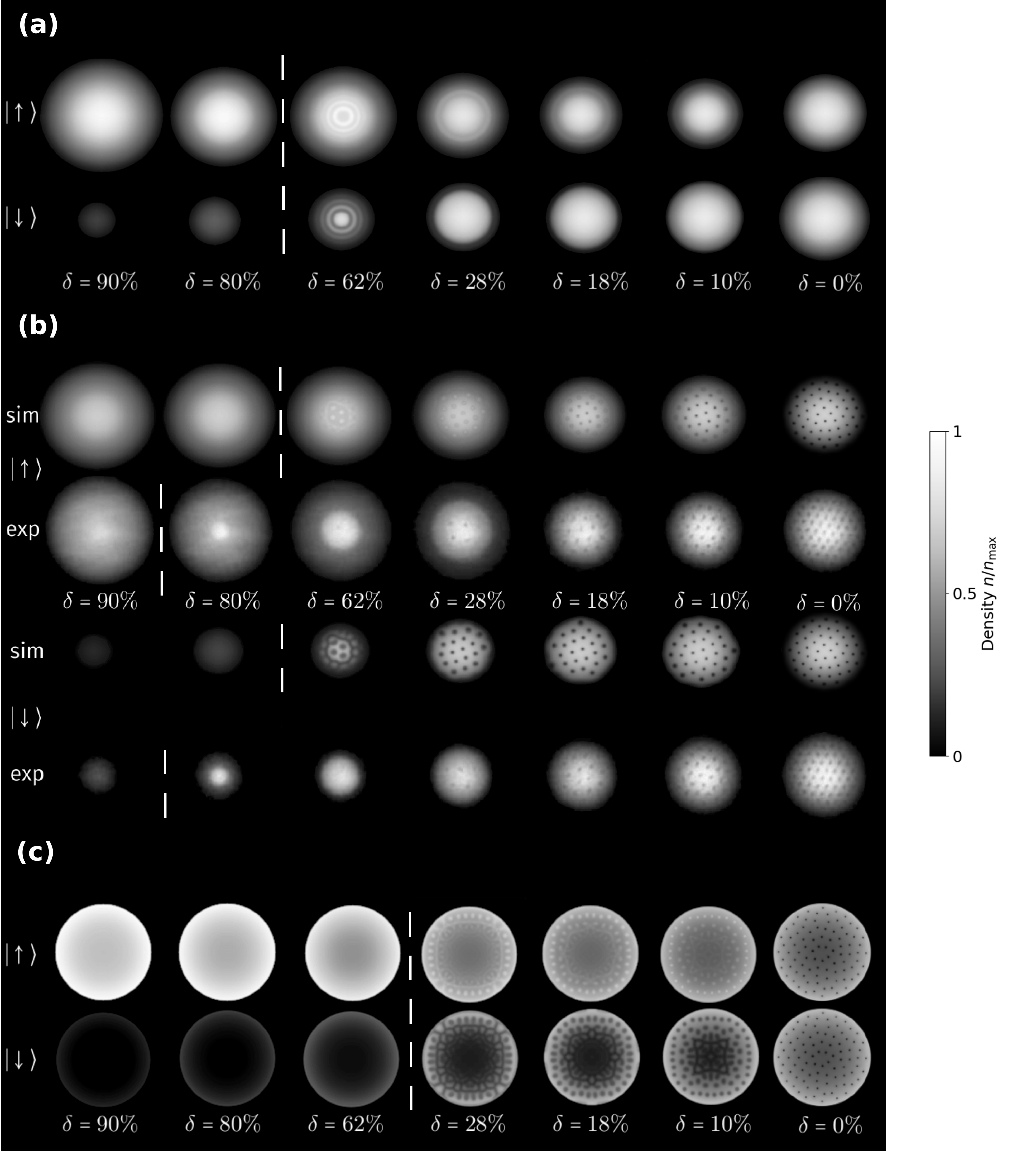}
    \caption{(a) Simulated densities $n(x,y, z = 0)$ for a non-rotating system, trapped in a harmonic potential, separately for the majority $|\uparrow \ \rangle$ and minority $|\downarrow \ \rangle$
    components and different population imbalances $\delta$. \\
    (b) Comparison between the experimental absorption images (exp) and the simulated densities $n(x,y, z=0)$ (sim) for a rotating system trapped in a harmonic potential, separately for the majority $|\uparrow \ \rangle$ and minority $|\downarrow \ \rangle$
    components and different population imbalances $\delta$. Experimental figures taken from M. Zwierlein {\it et al.}, Science {\bf 311}, 492 (2006). Reprinted with permission from AAAS.\\
    (c) Simulated densities $n(x,y, z = 0)$ for a rotating system, trapped in a box potential, separately for the majority $|\uparrow \ \rangle$ and minority $|\downarrow \ \rangle$
    components and different population imbalances $\delta$.\\
    The dashed line sets the cases with (on the right side of it) apart from the ones without (on the left) the superfluid phase\footnote{In the part (b), we can see the discrepancy in the position of this line as the experimental absorption image was taken for  $\frac{1}{\kF a}=0.2$ (i.e. slightly on the BEC side).}.}
    \label{fig:comp_gray}
\end{figure*}

\section{Harmonic oscillator trap}

\subsection{Ground state properties}

Before we start to consider the system in the rotating frame, we analyze ground states as a functions of the spin polarization $\delta$ revealed by the ASLDA method. The simulated density profiles are shown in Fig.~\ref{fig:comp_gray}a. First of all, we observe the phase separation there. The system spontaneously splits into the~spin-polarized corona
and the superfluid core. Hereafter, we define the polarized region as a volume where $p(\bm{r}) > p_{\mathrm{crit}}=50\%$. We find that this definition reasonably well selects the region where $\Delta(\bm{r})\approx 0$. The advantage of using it is that it is insensitive to the presence of quantum vortices and other singularity lines, where $\Delta=0$ by definition. 
In Fig.~\ref{fig:comp_gray}a the light grey center means there is a condensate with a high particle density. The darker space corresponds to unpaired particles. In the minority component, it is clearly seen for which population imbalance the condensation occurs. For $\delta=80$ and $90\%$, the dense part is no longer visible, so~the condensation vanishes there.

\begin{figure}[t]
    \centering
    \includegraphics[width=\linewidth]{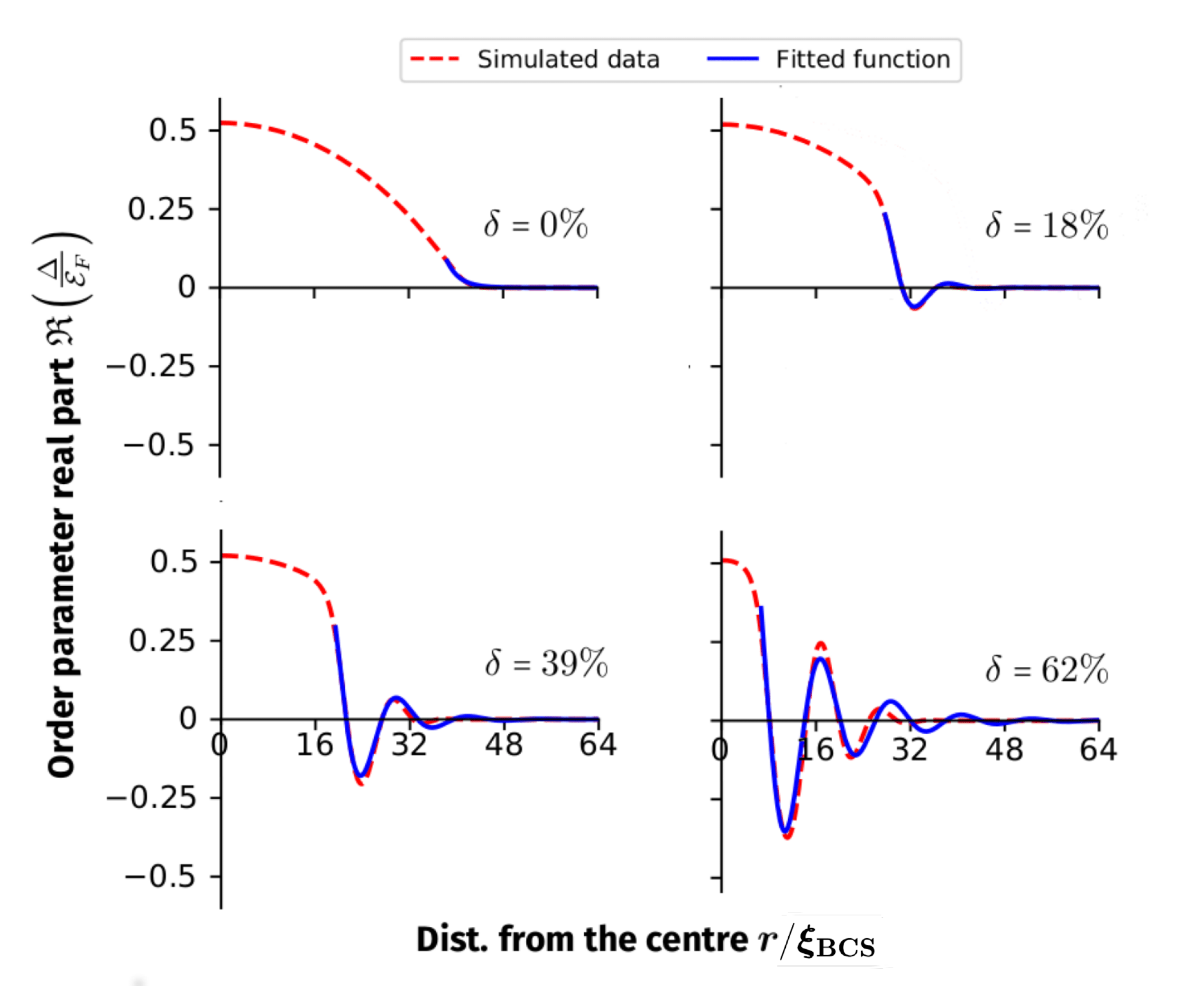}
    \caption{Real part of the order parameter $\Delta$ as a function of the distance from the system centre for different population imbalances - dashed (red) line. Fit to Eq.~(\ref{eq:Delta_osc}) is shown by solid (blue) line.}
    \label{fig:delta-sign}
\end{figure}

Our~attention should be drawn to the visible rings with a higher density (for~$\delta>0\%$, most clearly seen for $\delta=62\%$). The positions of the majority component density local maxima $n_\uparrow$ agree with the local minima positions of the minority component density $n_\downarrow$ and the other way round -- when $n_\uparrow$ reaches a~local minimum, $n_\downarrow$ takes a locally maximal value. The densities oscillate on the core-corona interface, and they are directly linked to oscillations of the order parameter $\Delta(\bm{r})$. The~number of oscillations ranges from $1$ to $5$ for the imbalances from $\delta=10\%$ to $62\%$. We can see that in Fig.~\ref{fig:delta-sign}. The order parameter vanishes for $\delta\gtrsim\delta_{\mathrm{crit}}\approx 80\%$. It is evidence that the fermionic condensate is no longer found there. The obtained $\delta_{\mathrm{crit}}$ agrees well with experimentally measured value, which is located in range $75-80\%$~\cite{Shin2006,Nascimbne2009,Navon2010,Olsen2015}. 
Contrary to the simulation, the oscillations of the density have not been detected so far experimentally. It needs to be emphasized that our geometry corresponds to the case of an extremely elongated trap in the $z$ direction. It is known that increasing the elongation of the trap also increases the strength of the oscillations~\cite{Pei2010,Kinnunen2018}. Thus, we expect that this particular feature is overestimated in our calculations in comparison to experimental realization. 

The oscillating order parameter is typically regarded as a signature of the FFLO state. However, proximity effects in superconductor-ferromagnet (SF) can also induce the oscillations. They are typically well described by \cite{Buzdin2005}:
\begin{equation}
\Delta(r)\propto \frac{1}{r-r_0}\exp\left(-\frac{r-r_0}{\xi_1}\right)\cos\left(\frac{r-r_0}{\xi_2}\right).
\label{eq:Delta_osc}
\end{equation}
The two characteristic length scales are correlation decay length $\xi_1$ and oscillating length $\xi_2$, $r_0$ is the location of the superfluid-ferromagnet transition.
Empirically, we find that 
the simulated data can be well fitted to the function~(\ref{eq:Delta_osc}), see~ Fig.~\ref{fig:delta-sign}.
For a spin-balanced system we set $\xi_2 \to \infty$. Then it is an analog of a superconductor-normal junction. It is expected to be the case, since at the edges of the cloud $T/\eF \gg 1$, and effectively the gas is in the normal state there. In case of the FFLO state, the oscillation length is expected to be related to mismatch of Fermi wave-vectors related to each spin-component $\xi_2^{-1}\sim k_{\textrm{F},\uparrow}-k_{\textrm{F},\downarrow}$, and should decrease as we increase $\delta$. Unfortunately, obtained value from the fit is fairly constant $k_\textrm{F}\xi_2\approx 2.77$.
For this reason, we conclude that for a non-rotating setup, phase separation is a dominant process. 

\subsection{System with rotation}

We start an analysis of results for the systems with rotation by comparing density sections $n_\sigma(x,y,z=0)$ with the absorption images taken experimentally~\cite{Zwierlein_2006}. This is done in Fig.~\ref{fig:comp_gray}b. It needs to be emphasized that the experimental measurements are done after ramping the magnetic field to the Bose-Einstein condensate (BEC) regime and releasing the gas from the trap. The target value of the magnetic field, rate of the ramping as well as expansion time after the release may impact the configuration in the final state~\cite{Matyjakiewicz2008,Schunck2007}. Moreover, the non-interacting gas in the corona expands differently from the interacting core~\cite{Shin2008}. Here we do not account for all these effects related to the experimental measurement --- we directly compare numerical density profiles at unitarity with experimental signal measured after sweeping to the molecular regime. We find that, in general, the ASLDA simulation results are qualitatively consistent with the experiment. It is quite remarkable since we do not make any attempts to match the experimental results (there are no free parameters in the theory that we fit to the experiment). Most likely, the rigidity of the order parameter with respect to the fast sweeps of the magnetic field is responsible for overall consistency between the experiment and the simulation~\cite{Szymaska2005}.
\begin{figure*}[t]
    \centering
    \includegraphics[width=.95\linewidth]{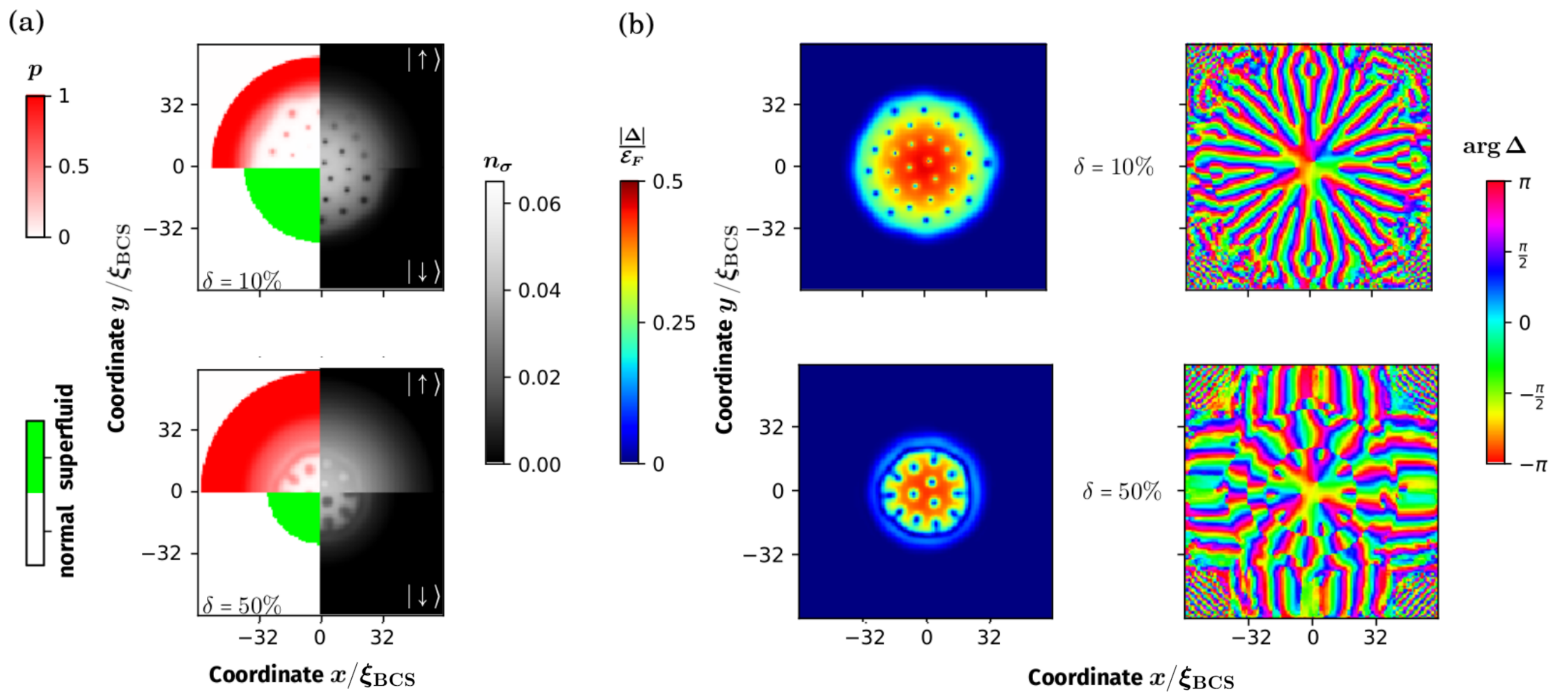}
    \caption{(a) Polarization $p(x,y,z = 0)$ (top left part of each subplot), densities $n_\sigma (x,y,z = 0)$ of the majority (top right part
    of each subplot) and the minority components (bottom right part of each subplot) as well as the indication of the superfluid region (bottom left part of each subplot). (b) Paring gap absolute value $|\Delta(x,y,z=0)|$ (left column) and its phase $\arg\Delta(x,y,z=0)$ (right column). The rows correspond to $\delta=10\%$ (top) and $50\%$ (bottom).}
    \label{fig:multiplot}
\end{figure*}

The ASLDA correctly reproduces the gross properties of the emerging vortex lattices (see also Fig.~\ref{fig:comp_vort}). For the population imbalances, where we see the superfluid fraction ($\delta$ from $0$ to $62\%$), the vortices appear. It is in agreement with the experiment where critical population imbalance for rotating case was reported as $\delta_c\approx70\%$. Some unpaired particles do~not leave the superfluid part and tend to occupy the vortices inside. It is visible when comparing the majority and minority component densities. We see that vortices (especially those outside the superfluid core) are filled with the spin up particles, while the opposite spin comment exhibit suppression in the core. It is due to the specific occupation of Andreev states localized in the vortex cores~\cite{Takahashi2006,Hu2007,2011.13021} Experimental images do not reveal such structure of the vortex cores. Instead, both spin components are suppressed in the core, irrespective of the population imbalance. To date, there are no theoretical studies that investigate the impact of the magnetic field sweep to the BEC side of the resonance on the vortex structure. Technically, this procedure is needed to visualize vortices, which otherwise are not detected directly at unitarity. The imaging operation (sweep and expansion) takes about $\Delta t_{\textrm{img.}}\sim 10^3\hbar/\eF$. On the other hand, the time needed by an atom at the Fermi surface to travel distance of the order of vortex core size is $\Delta t_{\xi}\sim 1\hbar/\eF$. Clearly, till imagining there is sufficient time for vortices to change their internal structure. Similar argumentation also applies to particles trapped in the vicinity of nodal lines associated with the FFLO state. In the case of slow ramps from the resonance to the molecular limit, it can be assumed that the process is adiabatic and then the results of static calculations as presented in~\cite{PhysRevB.99.134506} may be applicable. In such a case information about the initial vortex structure will be also lost. 

The detailed analysis of the vortex lattices in terms of the densities, polarization, and the superfluid core radii, for two selected population imbalances is shown in Fig.~\ref{fig:multiplot}. We can distinguish two types of regions that accumulate the spin polarization: vortex cores and, as for the case without rotation, the edge of the cloud.
This can  be  understood  based  on  energetic  considerations: unpaired particles tend to accumulate in regions where $\Delta\approx 0$ to do not break Cooper pairs in surrounding superfluid. The superfluid radii (volume  region where $\Delta>0$) are larger here than in the non-rotating case. On average, the radius increases by $16.5\%$ with the standard deviation of $4.8\%$. This enlargement is caused by centrifugal force. For the population imbalances $\delta=80$ and $90\%$, the superfluidity disappears.

For low population imbalances, vortices tend to organize into the triangular lattice in the superfluid core, while perturbations in the ordering emerge mainly close to the boundaries.
For $\delta=39,50$ and $62\%$, there are density oscillations also resulting in the oscillating polarization. They are correlated with the order parameter fluctuations, as in the case of the superfluid-ferromagnet proximity effect. Precisely, the oscillations are accompanied with additional jumps of phase by $\pi$, see case $\delta=50\%$ in Fig.~\ref{fig:multiplot}. It indicates that the FFLO-type phase may coexist with the vortex lattice for high population imbalances and, certainly, smaller than the critical population imbalance~$\delta_c$. 

\begin{figure}[h!]
    \centering
    \includegraphics[width=\linewidth]{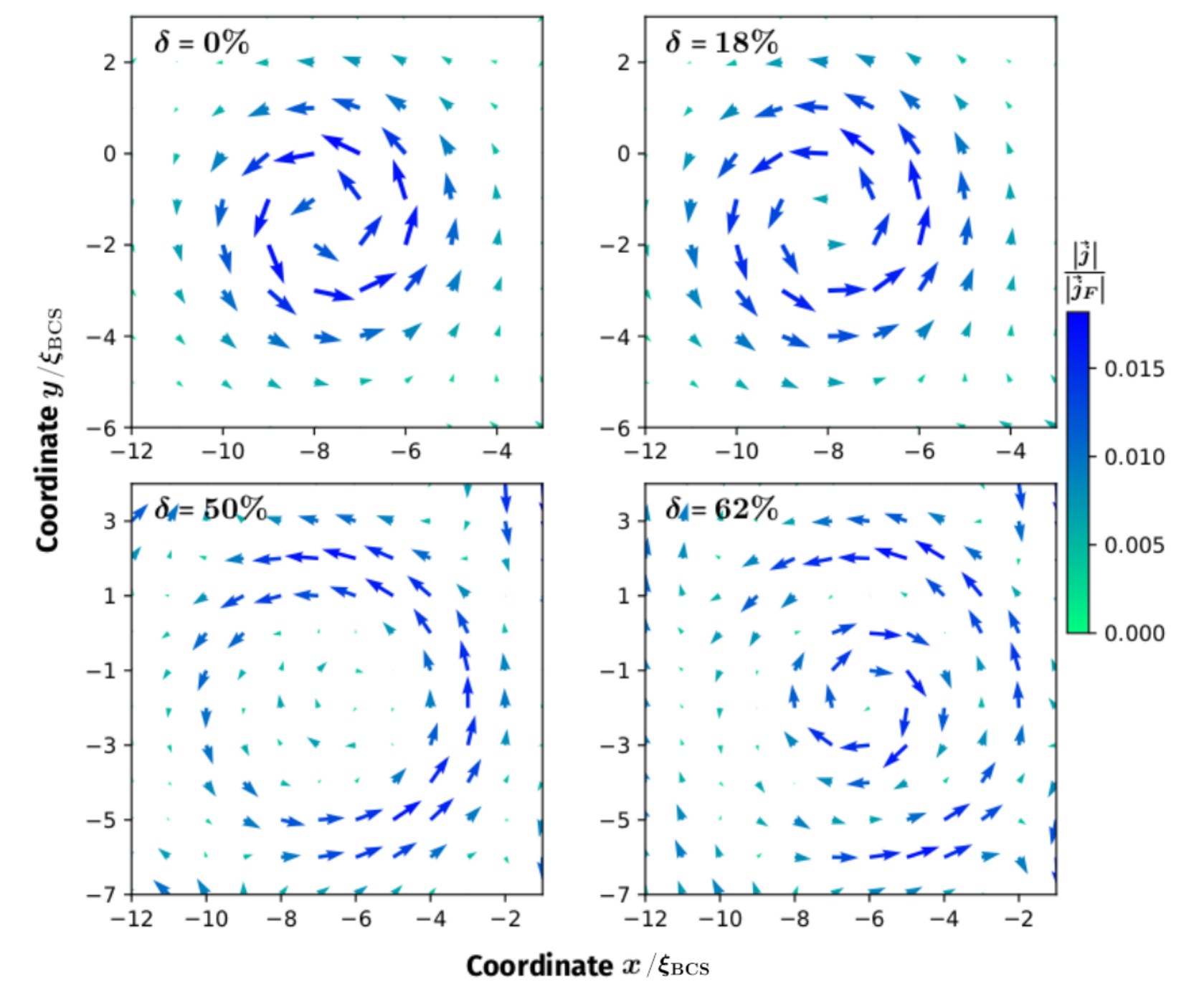}
    \caption{Probability current (total $\bm{j}^{\prime}=\bm{j}_{\uparrow}^{\prime}+\bm{j}_{\downarrow}^{\prime}$ for $\delta=0, 18, 62\%$, minority component $\bm{j}_{\downarrow}^{\prime}$ for $\delta=50\%$) in a rotating frame. Only the surroundings of selected vortices are presented. The reversed flow is visible for $\delta=50$ and $62\%$.}
    \label{fig:rev_currs}
\end{figure}

We have not identified any signatures  of quantum vortices hosting inside FFLO state, as suggested in~\cite{Inotani2021}. Instead, we find emergence of vortices with reversed circulation, predicted in~\cite{2011.13021}. To demonstrate this, we consider the current profile  $\bm{j}_{\sigma}(\bm{r})$. In general the current is dominated by overall rotation of the system with velocity $\bm{v}^{(\Omega)}(\bm{r})=\bm{\Omega}\times \bm{r}$, where $\bm{\Omega}=[0,0,\Omega_z]$ (see also Fig.~\ref{fig:velocities}). Thus, to extract current within rotating frame we apply transformation $\bm{j}_{\sigma}^{\prime}=\bm{j}_{\sigma}-n_\sigma \bm{v}^{(\Omega)}$. The current $\bm{j}_{\sigma}^{\prime}$ acquires significant values only in vicinity of quantum vortices. In Fig.~\ref{fig:rev_currs} we show it for representative vortices for a few selected population imbalances. 
In the systems with a small population imbalance such as $\delta=0$ or $18\%$, the total current in~the rotating frame is moving counter-clockwise around the vortex. This situation changes, when we increase the imbalance. In case of $\delta=50\%$ we may notice there is a small region inside the vortex, where the current is rotating clockwise. For the total current in the system with $\delta=62\%$ this effect is considerably more visible. Clearly, in the vortex lattices with high population imbalance we can find spin-polarized vortices with reversed flow.

\section{Impact of the trapping potential geometry}

\begin{figure}[h!]
    \centering
    \includegraphics[width=\linewidth]{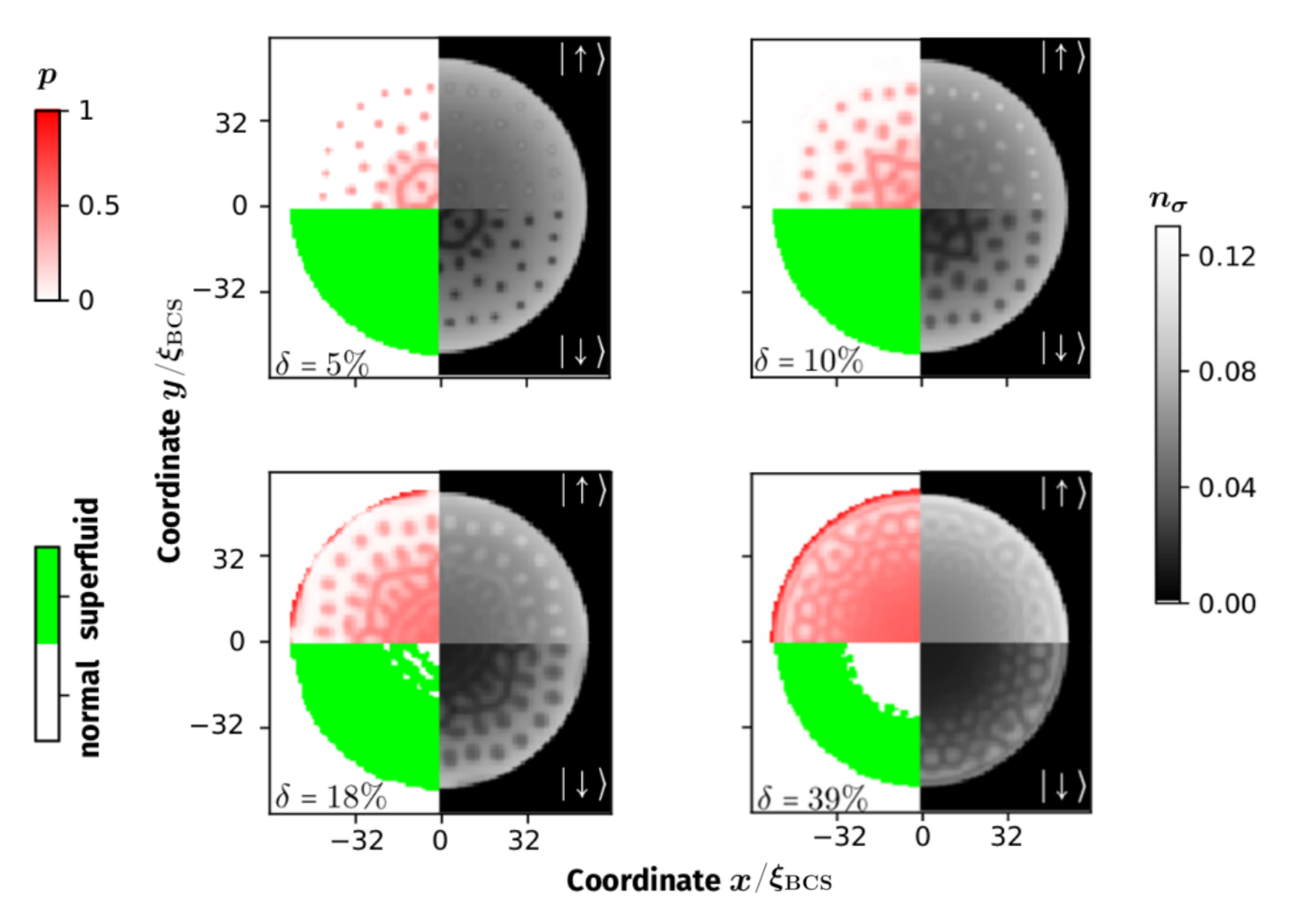}
    \caption{Polarization $p(x,y,z = 0)$ (top left part of each subplot), densities $n_\sigma (x,y,z = 0)$ of the majority (top right part
    of each subplot) and the minority components (bottom right part of each subplot) as well as the indication of the superfluid part (bottom left part of each subplot).}
    \label{fig:multiplot_tube}
\end{figure}

New generation of experiments with 3D ultracold gases utilize typically box-like traps~\cite{Gaunt2013,Chomaz2015,Navon2016,Mukherjee2017,2106.09716}. In this section we provide predictions of the ASLDA method in case of such a trap. The gas in the non-rotating box trap is uniform (except regions in close vicinity of boundaries). However, once we introduce rotation, we observe a fundamental change in the system. Neither the density maximum is placed in the system center, nor vortices create a triangular lattice, except the case $\delta=0\%$ where vortices tend to organize in the Abrikosov lattice. The lack of perfect simulated Abrikosov lattice for the spin symmetric system can be due to boundary effects; however, we cannot exclude that the
algorithm  stuck  in a metastable  stable  state  (although
we did a few tries with different starting points).  Due to the centrifugal force, the particles are expelled towards tube walls, and a structure resembling a meniscus is formed. It can be seen in Fig.~\ref{fig:comp_gray}c.
In such case, one can expect that the spin-imbalanced system separates into a spin-polarized core and a superfluid ring. Indeed, the unpaired particles occupy the less populated area and the vortex cores, see Fig.~\ref{fig:multiplot_tube}. One can also notice complex structures in the gas cloud formed by nodal lines. These are lines along which phase of the order parameter changes abruptly by $\pi$, and in Fig.~\ref{fig:multiplot_tube} they are seen especially well in polarization $p({\bm{r}})$ plots. While their presence may be a signature of the emergence of the FFLO-type phase, their complexity precludes stating that the resulting phase pattern can be approximated as a superposition of the vortex lattice phase and modulation of the sign along the radial direction. 
For example, in the cases $\delta=5\%$ and $10\%$ objects that emerge close to the tube center resemble recently predicted spin-polarized droplets~\cite{Magierski_2019} or soliton sacks~\cite{Barkman2020} that are filled by vortices. In Fig.~\ref{fig:multiplot_tube} their boundaries can be tracked by lines where the local polarization is non-zero. 
As we increase the imbalance above a threshold value, in the studied case above $\delta\gtrsim 18\%$, we start to destroy the superfluidity in the center. The system becomes phase-separated, and we have a spin-polarized core in its center. The order parameter has a close to zero value there and non-zero in the superfluid ring with an oscillatory behavior on the interface. The superfluidity vanishes entirely when $\delta = 62\%$.

\begin{figure}[h!]
    \centering
    \includegraphics[width=\linewidth]{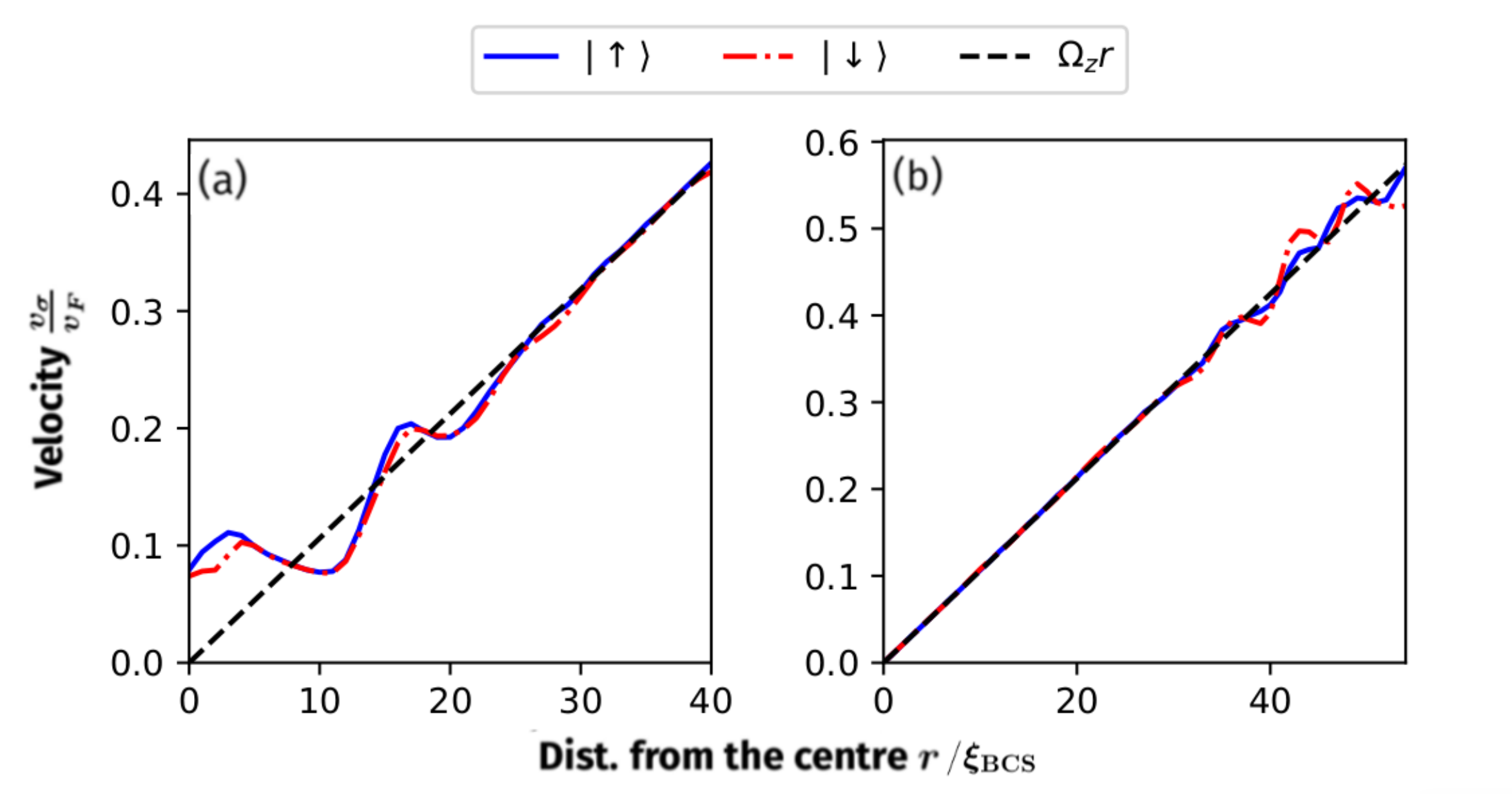}
    \caption{Velocity profiles of the majority $\left|\:\uparrow\:\right\rangle$ and minority $\left|\:\downarrow\:\right\rangle$ components as functions of the distance from the system centre $r$. Spin imbalance $\delta=39\%$. Trapping potential: (a) the harmonic oscillator and (b) the tube.}
    \label{fig:velocities}
\end{figure}
The velocity profiles of the flow provide further indication that the phase separation process is dominant for relatively large population imbalances. We expect the spin-polarized (normal) part to rotate like a solid body with $v(r)=\Omega_z r$. The superfluid part can only support rotation via quantum vortices, and the velocity profile should be different, though. It is shown in Fig.~\ref{fig:velocities}. In the gas trapped within the harmonic potential, the core is superfluid, so the velocity in this region differs from the one corresponding to the solid body rotation. As soon as we get to the corona, which is polarized, the profiles converge to expected $\Omega_z r$.
On the other hand, the system in the box trap has a spin-polarized core and a superfluid outer part. The velocity profile near the core is linear, and it changes in the superfluid ring.

Finally, let us compare the global properties of the vortex lattice. 
\begin{figure}[h!]
    \centering
    \includegraphics[width=0.75\linewidth]{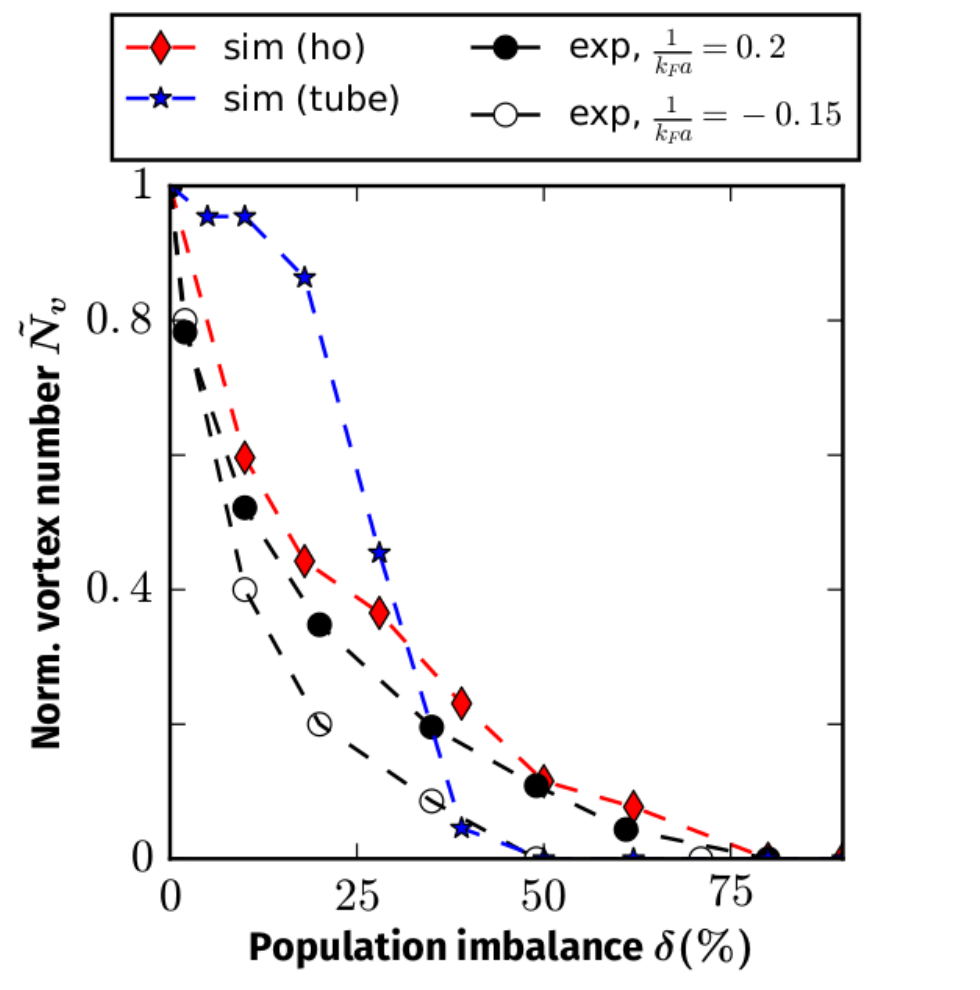}
    \caption{Normalized number of vortices $\tilde{N}_v/N_{v, \mathrm{max}}$ as a function of population
    imbalance $\delta$. Comparison between the experimental results (exp) for the Fermi gas on the BEC side ($\frac{1}{k_Fa} = 0.2$), the BCS side ($\frac{1}{k_Fa} = -0.15$) near the unitary regime and the simulation results (sim). Experimental  points are taken from~\cite{Zwierlein_2006}.}
\label{fig:comp_vort}
\end{figure}
In Fig.~\ref{fig:comp_vort} we present the normalized vortex numbers defined as $\tilde{N}_v(\delta)=\frac{N_{v}(\delta)}{N_{v,\mathrm{max}}}$ for different systems, including also experimental results for the harmonic trap. On qualitative level, experimental and simulated number of vortices as a function of the population  imbalance agrees well. The fact that the data series from simulation is slightly above the experimental data does not necessarily mean that the DFT model is incorrect at quantitative level. While in the simulation we can count number of vortices accurately, in the experiment vortices close to the cloud boundary are gradually lost during the expansion~\cite{Schunck2007}. That would lead to an underestimation of the number of vortices. Especially for higher population imbalances, where they are not easily distinguishable due to the presence of unpaired particles in the vortex cores.

The change of the external potential vastly affects the vortex number as a function of the population imbalance, see the result for the box trap. There is a sharp decrease in $\tilde{N}_v$ between the values $\delta=18$ and $39\%$. Up to $\delta=18\%$ almost the whole box is in a superfluid state, and above the spin-polarized region starts to develop in the center. At $\delta=39\%$, the remaining superfluid region close to the boundary becomes too small to accommodate any vortices. 

\section{Conclusions}
The results presented in this paper demonstrate that the properties of rotating and spin-imbalance ultracold fermionic gas strongly depend on the control parameters: confining potential and population imbalance. The spatial separation of the system into superfluid and normal components appears as the leading process. The precise location of these components is determined by the shape of the confining potential. The normal component is composed mainly of the atoms being in the same spin state and at the interference of the two phase proximity effects, typical for superconductor-ferromagnet junction, emerge. They include sign oscillations of the order parameter, an effect regarded as a signature of the FFLO state. Rotation induces additional singularity points, around which the phase rotates by $2\pi$, and the final pattern acquires a complex form. 
The presence of complex and non-triangular vortex patterns can be used as an indirect signature of the emergence of exotic superfluidity; however, this signal does not point directly to the FFLO phase.
In the context of searching of FFLO state, it is desired to minimize the number of inhomogenity sources and search for well-developed modulations of the order parameter over large scales, compared to coherence length. 

\begin{acknowledgments}
We thank Martin Zwierlein for the helpful insight into the experiment. We also thank Piotr Magierski for reading a draft of the
manuscript and for valuable criticism.
This work was supported by the Polish National Science Center (NCN) under
Contract No. UMO-2017/26/E/ST3/00428. In the final stage of work, JK was supported under Contract No. UMO-2019/34/E/ST2/00289.
We acknowledge Interdisciplinary Centre for Mathematical and Computational Modelling (ICM) of Warsaw University for computing resources at Okeanos (grant No. GA83-9) and PL-Grid Infrastructure for providing us resources at Prometheus supercomputer.

Center for Theoretical Physics of the Polish Academy of Sciences is a member of the National Laboratory of Atomic, Molecular and Optical Physics (KL FAMO).
\end{acknowledgments}

\bibliography{apsbiblio}
\end{document}